\documentclass[superscriptaddress,amsmath,amssymb,longbibliography,reprint,twocolumn]{revtex4-1}

\usepackage[english]{babel}
\PassOptionsToPackage{english}{babel}
\usepackage{lineno}
\usepackage[utf8]{inputenc}
\usepackage{color}
\usepackage{siunitx}
\usepackage{graphicx,xcolor,hyperref}
\usepackage{booktabs}
\usepackage{soul}

\usepackage{hyperref}
\usepackage{makecell}
\hypersetup{
    bookmarks=false,         
    unicode=false,          
    pdftoolbar=false,        
    pdfmenubar=true,        
    pdffitwindow=false,     
    pdfstartview={FitH},    
    pdftitle={Gate-tunable, superconductor-semiconductor parametric amplifier},    
    pdfauthor={D. Phan et al},     
    pdfsubject={},   
    pdfcreator={},   
    pdfproducer={}, 
    pdfkeywords={}, 
    pdfnewwindow=true,      
    colorlinks=true,       
    linkcolor=black,          
    citecolor=blue,        
    filecolor=magenta,      
    urlcolor=blue           
}

\newcommand{\sinc}{\mathop{\mathrm{sinc}}}

\begin{document}
\sisetup{tight-spacing=true}

\selectlanguage{english}
\title{Gate-tunable, superconductor-semiconductor parametric amplifier}

\author{D. Phan}
\affiliation{IST Austria, Am Campus 1, 3400 Klosterneuburg, Austria}
\author{P. Falthansl-Scheinecker}
\affiliation{IST Austria, Am Campus 1, 3400 Klosterneuburg, Austria}
\author{U. Mishra}
\affiliation{IST Austria, Am Campus 1, 3400 Klosterneuburg, Austria}
\author{W.M. Strickland}
\affiliation{Center for Quantum Information Physics, Department of Physics, New York University, New York, NY, 10003, USA}
\author{D. Langone}
\affiliation{Center for Quantum Information Physics, Department of Physics, New York University, New York, NY, 10003, USA}
\author{J. Shabani}
\affiliation{Center for Quantum Information Physics, Department of Physics, New York University, New York, NY, 10003, USA}
\email{js10080@nyu.edu}
\author{A.P. Higginbotham}
\affiliation{IST Austria, Am Campus 1, 3400 Klosterneuburg, Austria}
\email{andrew.higginbotham@ist.ac.at}

\date{\today}

\begin{abstract}
We have built a parametric amplifier with a Josephson field effect transistor (JoFET) as the active element.
The device's resonant frequency is field-effect tunable over a range of $2~\mathrm{GHz}$.
The JoFET amplifier has $20~\mathrm{dB}$ of gain, $4~\mathrm{MHz}$ of instantaneous bandwidth, and a  $1~\mathrm{dB}$ compression point of $-125.5~\mathrm{dBm}$ when operated at a fixed resonance frequency.
\end{abstract}

\maketitle

\section{Introduction}
Quantum-limited amplifiers are, for many experimental platforms, the first link in the quantum signal processing chain, allowing minute signals to be measured by noisy, classical electronics \cite{haus_quantum_1962,caves_quantum_1982,clerk_introduction_2010}.
Whereas later parts of the chain are dominated by semiconductor-based devices, the quantum-limited step is usually performed using metallic superconductors \cite{yurke_observation_1989,movshovich_observation_1990,yurke_low-noise_1996,castellanos-beltran_widely_2007,castellanos-beltran_amplification_2008,bergeal_phase-preserving_2010,macklin_near-quantum-limited_2015,planat_photonic-crystal_2020,fratitini_optimizing_2018}, although very recently quantum-limited amplifier have been demonstrated using graphene weak links \cite{butseraen_gate-tunable_2022,sarkar_quantum_2022}.
Aluminum-oxide based tunnel junctions have proven to be more reliable and stable than any other platform, thanks to the formation of pristine Al-AlO$_x$ interfaces through the natural oxidization of Al.

A comparable natural, coherent, and scalable interface between a superconductor and a semiconductor was only recently introduced in the form of Al-InAs hybrid two-dimensional electron gas heterostructures \cite{shabani_two-dimensional_2016,sarney_reactivity_2018, sarney_aluminum_2020}.
Josephson junctions fabricated on these materials yield a voltage-controllable supercurrent with highly transparent contacts between Al and InAs quantum wells \cite{kjaergaard_quantized_2016,mayer_superconducting_2019,dartiailh_missing_2021}.
Al-InAs has recently been instrumental in exploring topological superconductivity \cite{nichele_scaling_2017,fornieri_evidence_2019,dartiailh_phase_2021,aghaee_inas-al_2022}, mesoscopic superconductivity \cite{bottcher_superconducting_2018}, and voltage-tunable superconducting qubits \cite{casparis_superconducting_2018}. 
Al-InAs hybrids have also been used to demonstrate magnetic-field compatible superconducting resonators \cite{phan_detecting_2022} and qubits \cite{pita-vidal_gate-tunable_2020,kringhoj_magnetic-field-compatible_2021}.
There has also been impressive progress on realizing all-metallic field compatible parametric amplifiers recently \cite{xu_magnetic_2022,khalifa_nonlinearity_2022,vine_in-situ_2022}.

More broadly, the inherent scalability of semiconductors has motivated a great deal of research on quantum applications, including the scalable generation of quantum-control signals \cite{schaal_cmos_2019,pauka_cryogenic_2021,xue_cmos-based_2021,lecocq_control_2021}, amplification \cite{cochrane_parametric_2022}, and the processing of quantum information \cite{larsen_semiconductor-nanowire-based_2015,casparis_superconducting_2018} at fault-tolerant thresholds \cite{veldhorst_addressable_2014,takeda_fault-tolerant_2016,yoneda_quantum-dot_2018,noiri_fast_2022,xue_computing_2022}.

Here, we introduce Al-InAs as the basis for quantum signal processing devices.
We demonstrate a gate-tunable parametric amplifier using an Al-InAs Josephson field-effect transistor (JoFET) as the active element.
Our device has a resonant frequency tunable over 2 GHz via the field effect.
In optimal operating ranges, the JoFET amplifier has 20 dB of gain with a 4 MHz instantaneous bandwidth.
The gain is sufficient for integration into a measurement chain with conventional semiconductor amplifiers.
Accordingly, we find that the amplifier dramatically improves signal recovery when used at the beginning of a typical measurement chain.
We quantify noise performance by calibrating our classical measurement chain, measuring the insertion loss of all components used to connect the chain to the JoFET amplifier, and then referring measured noise to the JoFET input.
This procedure suggests that the JoFET amplifier's input-referred noise approaches the limits imposed by quantum mechanics, although a calibrated noise source at device input is needed to definitively verify quantum-limited performance.
Motivated by the successful operation of Al-InAs hybrid microwave circuits at large magnetic fields \cite{pita-vidal_gate-tunable_2020,kringhoj_magnetic-field-compatible_2021,phan_detecting_2022}, magnetic field compatibility is investigated.
Performance is noticeably degraded by even small $15~\mathrm{mT}$ parallel magnetic fields.
In contrast to metallic superconducting amplifiers, our platform is gate-tunable, and is natural to use as a detector for voltages from a high-impedance source.

\begin{figure}[bt]
	\centering
	  \includegraphics[scale=1]{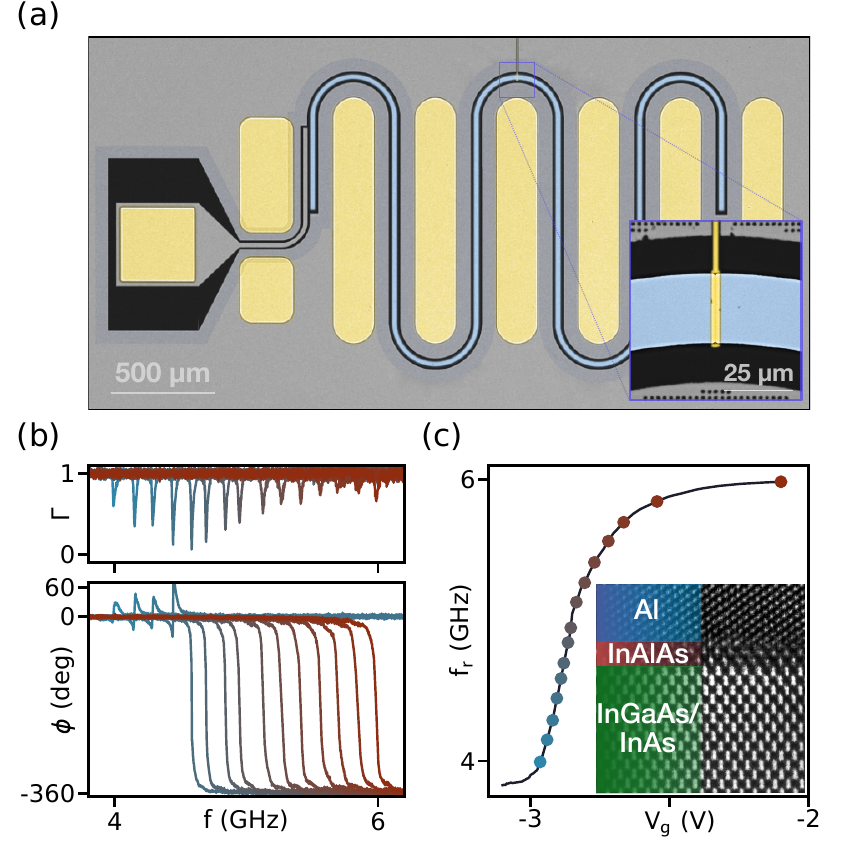}
		\caption{
		Device and frequency modulation via the field effect.
		(a) False-colored image of the device (InP substrate dark gray, Nb light gray, aluminum light blue, Au yellow). A feed line (left) capacitively couples to the resonator, which is formed from an Al-InAs semiconductor heterostructure. 
		Inset: Zoom of JoFET and electrostatic gate. 
		(b) Reflected signal magnitude $\Gamma$ (top) and phase $\phi$ (bottom) versus signal frequency $f$ at 15 different gate voltages sampled at equally spaced resonant frequency from $4~\mathrm{GHz}$ to $6~\mathrm{GHz}$. 
		Trace colors correspond to point colors in (c).
		This shows the shift of resonant frequency with decreasing (more negative) gate voltage.
		(c) Resonant frequency $f_r$ as a function of gate voltage $V_g$.
		Dot colors correspond to trace colors in (b).
		Inset: Cross-sectional transmission electron micrograph at the interface of Al thin film and InAs quantum well.
		The contrast and brightness have been enhanced for clarity.
		}
	  \label{fig:1}
\end{figure}

\section{Device design}
The JoFET amplifier is implemented as a half-wave coplanar waveguide (CPW) resonator with a gated superconductor-semiconductor hybrid Josephson field-effect-transistor positioned at the voltage node [Fig.~\ref{fig:1}(a)]. 
Ground planes of the device are formed from $50~\mathrm{nm}$ thin film Nb with $1~\mathrm{\mu m}$-squared flux-pinning holes near the edges to improve magnetic-field resilience \cite{samkharadze_high-kinetic-inductance_2016,kroll_magnetic-field-resilient_2019}.
The entire center pin of the resonator is made of Al-InAs heterostructure, defined by chemical etch.
The JoFET is defined by selective removal of the Al, followed by atomic layer deposition of alumina, and finally an electrostatic gate is defined with electron beam lithography and Au evaporation.
Large Au chip-to-chip bond pads are also co-deposited in the final step to improve wire-bonding yield.
The characteristic impedance of the CPW is designed to be near $50~\mathrm{\Omega}$ with $25~\mathrm{\mu m}$ wide center conductor width and a $16~\mathrm{\mu m}$ gap to the ground plane.
The device is capacitively coupled to an open $50~\mathrm{\Omega}$ transmission line, which forms the only measurement port.

We targeted a geometric resonant frequency of approximately $6~\mathrm{GHz}$ for compatibility with the standard $4-8~\mathrm{GHz}$ band used in circuit quantum electrodynamics experiments.
In designing the circuit, it is important to account for the kinetic inductance of the Al-InAs heterostructure, which in previous work caused a 20\% reduction in the circuit resonant frequency \cite{phan_detecting_2022}.
The device is therefore designed with a geometric resonant frequency of $7.2~\mathrm{GHz}$, corresponding to a CPW resonator length of $8~\mathrm{mm}$.
We have not directly measured the kinetic contribution for this heterostructure growth, although we note that the Al is slightly thicker than the growth used in Ref.~\cite{phan_detecting_2022} which should result in a relatively higher resonant frequency for the devices in this work.

The external coupling $\kappa_\mathrm{ex}$ and the critical current $I_c$ determine the operating bandwidth and dynamic range of the amplifier \cite{eichler_controlling_2014}.
While $\kappa_\mathrm{ex}$ can be estimated from the circuit geometry, $I_c$ is more subtle because it depends on material details.
Based on independent transport tests, we found that a JoFET with a width of $25~\mathrm{\mu m}$ have an expected critical current of $10~\mathrm{\mu A}$ at positive gate voltages.
During design, we used Ref.~\cite{eichler_controlling_2014} to estimate that this critical current should yield a Kerr nonlinearity on the order of a few kHz.
A more detailed discussion of the Kerr nonlinearity for our geometry and its relationship to critical current is given in Sec.~\ref{sec:nl_and_amp}.
Based on the dynamic range calculation in Ref.~\cite{eichler_controlling_2014}, a $1~\mathrm{dB}$ compression point of a few photons is expected when operating with $20~\mathrm{dB}$ of gain.

\section{Gate tunability}
The complex microwave reflection coefficient $R=\Gamma e^{i \phi}$ of a small incident signal, with $\Gamma$ being the magnitude and $\phi$ being the phase, is measured from the sample in a dilution refrigerator with a standard measurement chain, including a cryogenic commercial high-electron mobility transistor amplifier (HEMT).
For large (more positive) gate voltages, the measured reflection coefficient displays a small dip in magnitude and a $360^{\circ}$ winding of phase, signaling that the resonator is strongly coupled to the measurement port and that dissipation is weak [Fig.~\ref{fig:1}(b)].
Indeed, fitting the reflection coefficient to a one-port model gives an external coupling efficiency $\kappa_\mathrm{ex}/\kappa_\mathrm{tot} = 0.83$ at zero gate voltage.

Application of negative gate voltage results in dramatic changes in resonator frequency [Fig.~\ref{fig:1}(b)], indicating that the JoFET contributes a gate-tunable inductance to the resonator.
The resonant frequency can be tuned over more than $2~\mathrm{GHz}$ of bandwidth [Fig.~\ref{fig:1}(c)], exhibiting weak voltage dependence at extremal values, reminiscent of a transistor reaching pinch-off at negative voltage and saturation at positive voltage.

\begin{figure}[bt]
	\centering
	  \includegraphics[scale=1]{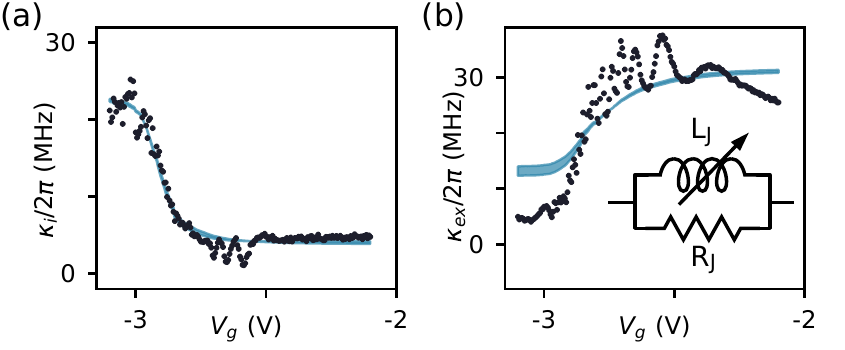}
		\caption{ Circuit model. 
		Internal loss rate $\kappa_i$ (a) and external coupling $\kappa_\mathrm{ex}$ (b) versus gate voltage $V_g$.
		Experimental data indicated by black points, and results of parallel $RL$ model calculated for a range of bare resonant frequencies, $f_0=6.00-6.45~\mathrm{GHz}$, indicated by blue shaded region.
		Inset: Proposed JoFET linear-equivalent circuit, consisting of tunable inductance $L_J$ and fixed shunt resistance $R_J$ in parallel.
		}
		\label{fig:2}
\end{figure}

The internal dissipation rate $\kappa_i$ and external coupling rates $\kappa_\mathrm{ex}$ also evolve with gate voltage.
Near JoFET pinch off $\kappa_i$ increases sharply, and $\kappa_\mathrm{ex}$ decreases over the same range [Fig.~\ref{fig:2}(a),\ref{fig:2}(b)].
The net effect is therefore a decrease in coupling efficiency near pinch-off, which limits the usable frequency range of the JoFET amplifier.

A plausible origin of the increase in $\kappa_i$ is dissipation in the JoFET region.
To test this hypothesis, we consider a minimal model of the JoFET as a parallel resistor $R_J$ and inductor $L_J$ [Fig.~\ref{fig:2}(b),inset], and introduce an effective resistance $R_J/(\Delta \overline{u})^2$ where $\Delta \overline{u}$ is the normalized flux drop across the resistor.
The Josephson inductance and flux drop can be inferred from the measured resonant frequency if the bare circuit parameters without the Josephson junction are known (see Appendix \ref{app:rlc}).
In our case the bare circuit resonance is not precisely known because we have not characterized the heterostructure kinetic inductance for this particular material growth.
We therefore model a range of possible bare resonant frequencies $f_0$.
The lower value of $6~\mathrm{GHz}$ is taken from the resonant frequency at high gate voltage, and the upper value of $6.45~\mathrm{GHz}$ is chosen to qualitatively match our measured Kerr nonlinearities, to be discussed later.
The high-range value corresponds to a 10\% reduction from the designed geometric resonance of $7.2~\mathrm{GHz}$, which, as one would expect, would imply that the current heterostructures have less kinetic inductance than those in our earlier studies on structures with thinner Al \cite{phan_detecting_2022}.

Fitting $\kappa_i$ to the circuit model results in agreement with the data, with a best-fit shunt resistance $R_J =15 \pm 2~\mathrm{k \Omega}$, where the uncertainty is derived from the possible range in $f_0$.
Based on this agreement, we conclude that the junction indeed introduces dissipation to the circuit, with the practical effect of limiting performance at high inductances.
The microscopic origin of this dissipation is not understood.
Candidate explanations are coupling to a lossy parasitic mode associated with the normal-conducting (Au) electrostatic gate, or a mechanism that is intrinsic to the Al-InAs material system.
Turning now to external coupling, the simple RLC model captures gate dependence of $\kappa_\mathrm{ex}$ at a qualitative level [Fig.~\ref{fig:2}(b)], although observed gate-dependence quantitatively exceeds theoretical expectations.
This discrepancy may reflect the role of extra capacitance in the JoFET region introduced by the electrostatic gate, which is not accounted for in the model.

\begin{figure}[bt]
	\centering
	  \includegraphics[scale=1]{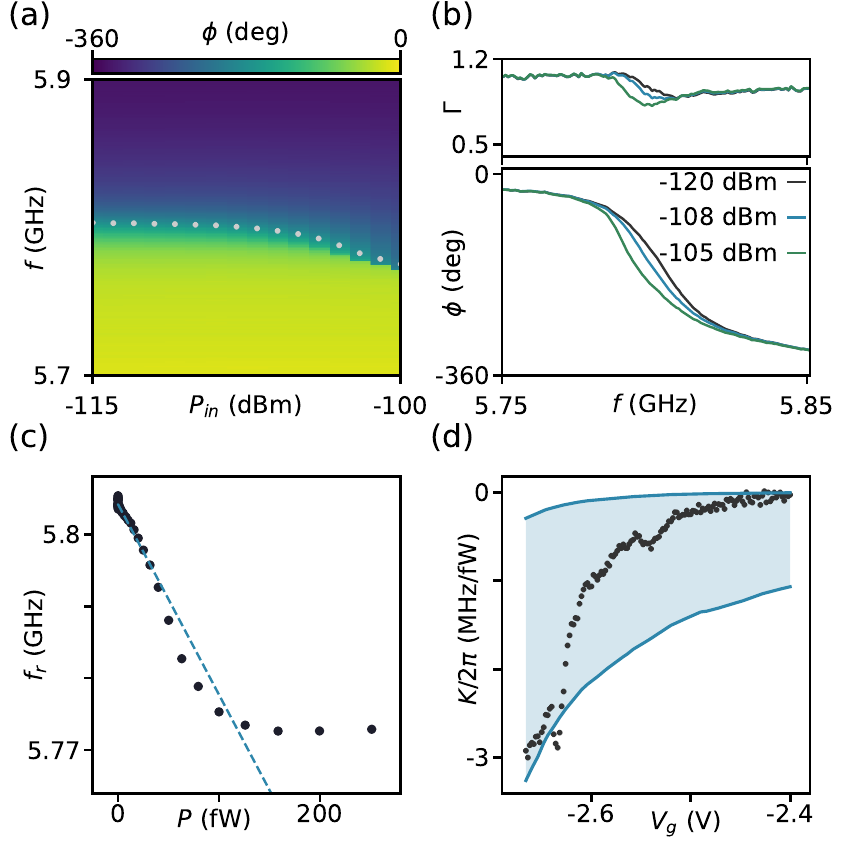}
		\caption{ Nonlinear response.
		(a) Reflected signal phase $\phi$ versus power at resonator input $P_\mathrm{in}$ and frequency $f$.
		White dots indicate resonant frequencies $f_r$.
		(b) Reflected magnitude $\Gamma$ (top) and phase $\phi$ (bottom) versus $f$ at different input powers: low (black), medium (blue), and close to critical power (green).
		(c) Estimation of Kerr nonlinearity: resonant frequencies extracted from (a) versus average network analyzer power applied at room temperature (dots). Slope of linear fit (dashed line) gives an estimate of the Kerr coefficient.
		(d) Measured gate-dependent Kerr coefficient (black dots), and expected value from the parallel RL model for a range of bare resonance frequencies $f_0=6.00-6.45~\mathrm{GHz}$ (blue region).
		Kerr coefficient is expressed as a frequency shift per power from the VNA at resonator input.
		Different dataset from (c).
		}
		\label{fig:3}
\end{figure}

\section{Nonlinearities and amplification}
\label{sec:nl_and_amp}
In addition to tunable inductance, the presence of the JoFET imparts a power-driven nonlinearity to the resonator \cite{eichler_controlling_2014}.
Measuring the reflected phase as a function of signal frequency and power reveals that the resonant frequency smoothly decreases with increasing input power $P_\mathrm{in}$ [Fig.~\ref{fig:3}(a)].
The output power from microwave sources is related to the resonator input power by an estimated total attenuation of $110~\mathrm{dB}$ for all measurements.
The downward shift in resonant frequency is accompanied by ``sharpening'' of the phase response [Fig.~\ref{fig:3}(b)].
Downward shift in resonant frequency and an alteration in lineshape are key qualitative signatures of the required Kerr nonlinearity $K$, which is useful for parametric amplification.
The Kerr nonlinearity is estimated by measuring the change in resonant frequency per incident power in the low power limit, as shown by the linear fit in Fig.~\ref{fig:3}(c).
Following this procedure at different gate voltages reveals that the nonlinearity is tunable with voltage, increasing in magnitude as gate voltage is decreased [Fig.~\ref{fig:3}(d)], qualitatively mirroring the decrease in circuit resonant frequency observed in Fig.~\ref{fig:1}(c).
To make the relationship between the observed Kerr nonlinearity and the circuit parameters more concrete, we have estimated the expected Kerr nonlinearity based on our circuit model, using the formula for a low-transmission Josephson junction (shown in Fig.~\ref{fig:3}(d), see Appendix~\ref{app:kerr} for details).
Unlike the case of linear circuit parameters, the uncertainty in the bare circuit resonance frequency ($f_0=6.00-6.45~\mathrm{GHz}$) translates into many orders of magnitude of uncertainty in the Kerr nonlinearity, reflecting the fact that the Kerr nonlinearity is a fourth-order effect.
Despite these challenges, our estimate indicates that the magnitude of the Kerr nonlinearity we observe is compatible with the circuit model considering this range of resonant frequencies.
It is important to note that the formula that we compare with is valid only for low-transparency Josephson junctions, which is not the case for our system \cite{kjaergaard_transparent_2017,mayer_superconducting_2019}.
If the bare circuit resonance were accurately known, a more complete treatment could possibly infer the true junction transparency based on the observed Kerr nonlinearity.

\begin{figure}[bt]
	\centering
	  \includegraphics[scale=1]{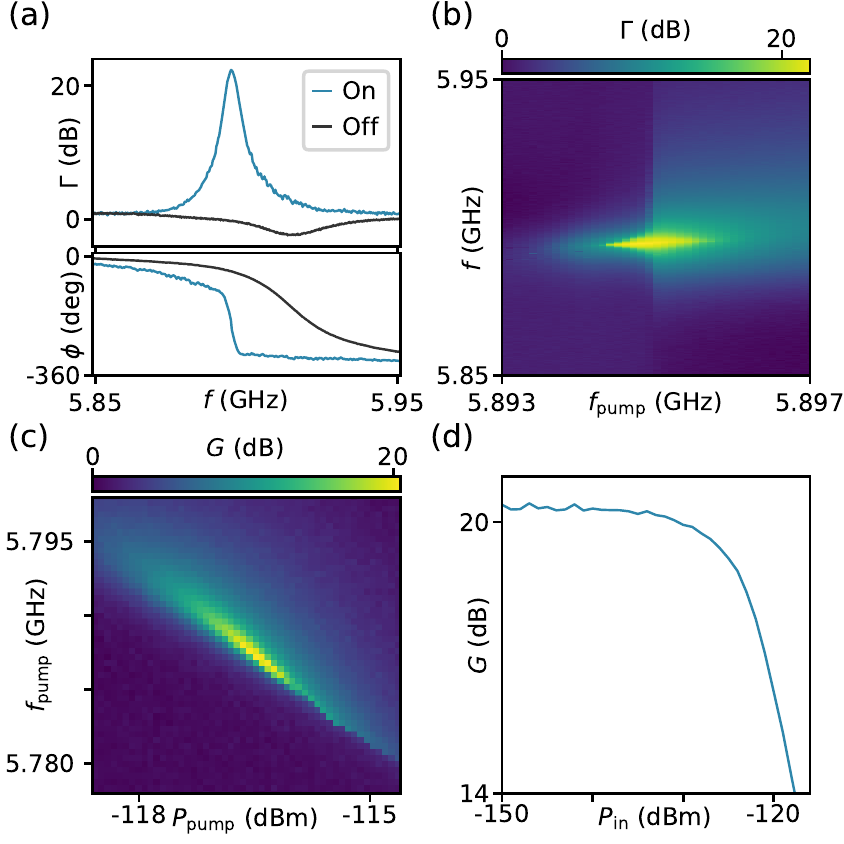}
		\caption{
		Parametric gain.
		(a) Reflected signal magnitude $\Gamma$ and phase $\phi$ measured as a function of signal frequency $f$ from bare resonator (black) and pumped resonator (blue).
		(b) $\Gamma$ measured as a function of pump frequency $f_\mathrm{pump}$ and signal frequency $f$.
		The pump power is fixed slightly below the critical pump power at which the system bifurcates.
		$V_g=-2.5~\mathrm{V}$ for (a)-(b).
		(c) Gain $G$ inferred from the maximal reflected amplitude measured as a function of pump power $P_\mathrm{pump}$ and pump frequency $f_\mathrm{pump}$, revealing line of maximum gain.
		$V_g=-2.55~\mathrm{V}$.
		(d) Gain $G$ measured as a function of signal power at device input $P_\mathrm{in}$.
		$V_g=-2.35~\mathrm{V}$, $f_{pump}=5.942~\mathrm{GHz}$, signal-pump-detuning $\delta f=1~\mathrm{MHz}$ and remains within the $3.6~\mathrm{MHz}$ instantaneous amplification bandwidth.
		}
		\label{fig:4}
\end{figure}

Parametric amplification is generated by applying a strong pump tone red detuned from the bare resonant frequency, in the vicinity of the phase ``sharpening'' features already identified close to the critical power [Fig.~\ref{fig:3}(b)].
Measuring scattering parameters with a weak probe signal reveals in excess of 20 dB of gain with a 4 MHz bandwidth, and a sharp phase response typical of a parametric amplifier [Fig.~\ref{fig:4}(a),\ref{fig:4}(b)].
Modest detunings in the pump frequency do not substantially affect the amplifier response, but for large deviations the gain decreases towards unity [Fig.~\ref{fig:4}(b)].
Measuring the amplifier gain $G$ as a function of pump frequency $f_\mathrm{pump}$ and input power $P_\mathrm{pump}$ reveals a continuous region of maximum gain with an easily identifiable optimum operating region, as expected for a parametric amplifier [Fig.~\ref{fig:4}(c)].
The gate voltages used for the gain optimization and measurement shown in Fig.~\ref{fig:4}(a),\ref{fig:4}(b) and in Fig.~\ref{fig:4}(c),\ref{fig:4}(d) are different due to gate instability, hysteresis, and shifts between cooldowns.
These particular datasets were taken weeks apart, separated by many wide-range gate voltage sweeps, and in different cooldowns.
Typically, after changing the gate voltage by a large amount (on the scale of Volts) it takes 15 minutes for the resonance to stabilize, and we find that there is hysteresis after large sweeps.
In general the resonant frequency remains stable for circa an hour which is sufficient for our gain and noise measurements.
After approximately one hour we often found measurable changes in amplifier gain and noise performance due to resonant frequency drifts, which could be compensated by re-optimization of the pump frequency and amplitude.
We emphasize that the drifts in the resonant frequency over this time period are much smaller than the linewidth, but still noticeable when operating at high gain.
We anticipate that optimization of the gate dielectric can greatly improve stability in future devices.

The power handling capability of the amplifier is quantified by measuring the gain for different signal powers [Fig.~\ref{fig:4}(d)].
At very low input signal power the gain saturates at $20.3~\mathrm{dB}$ [Fig.~\ref{fig:4}(d)].
Large input powers cause the amplifier gain to decrease, giving a $1~\mathrm{dB}$ compression point at input power of $-125.5~\mathrm{dBm}$ [Fig.~\ref{fig:4}(d)].
The gain and instantaneous bandwidth are comparable to early, parametric amplifiers based on metallic Al/AlOx Josephson junctions \cite{castellanos-beltran_widely_2007,castellanos-beltran_amplification_2008,bergeal_analog_2010}.
The frequency tunability of our resonator is also comparable to early parametric amplifiers \cite{castellanos-beltran_widely_2007}, although we have not demonstrated tunable amplification, which would currently be limited to the low-loss frequencies (see Figs. \ref{fig:1}-\ref{fig:2}).
The 1 dB compression point is only slightly lower than some early parametric amplifiers (two times below Ref.~\cite{zhou_high-gain_2014}), but orders of magnitude below modern implementations \cite{macklin_near-quantum-limited_2015,planat_photonic-crystal_2020}.
Both the resonator loss and the compression point need to be significantly improved for this amplifier to be practically useful.
In contrast to a recently-demonstrated semiconductor parametric amplifier with $3~\mathrm{dB}$ of gain \cite{cochrane_parametric_2022}, our device has sufficient gain ($20~\mathrm{dB}$) to overwhelm the noise of later parts of the measurement chain.

\begin{figure}[bt]
	\centering
	  \includegraphics[scale=1]{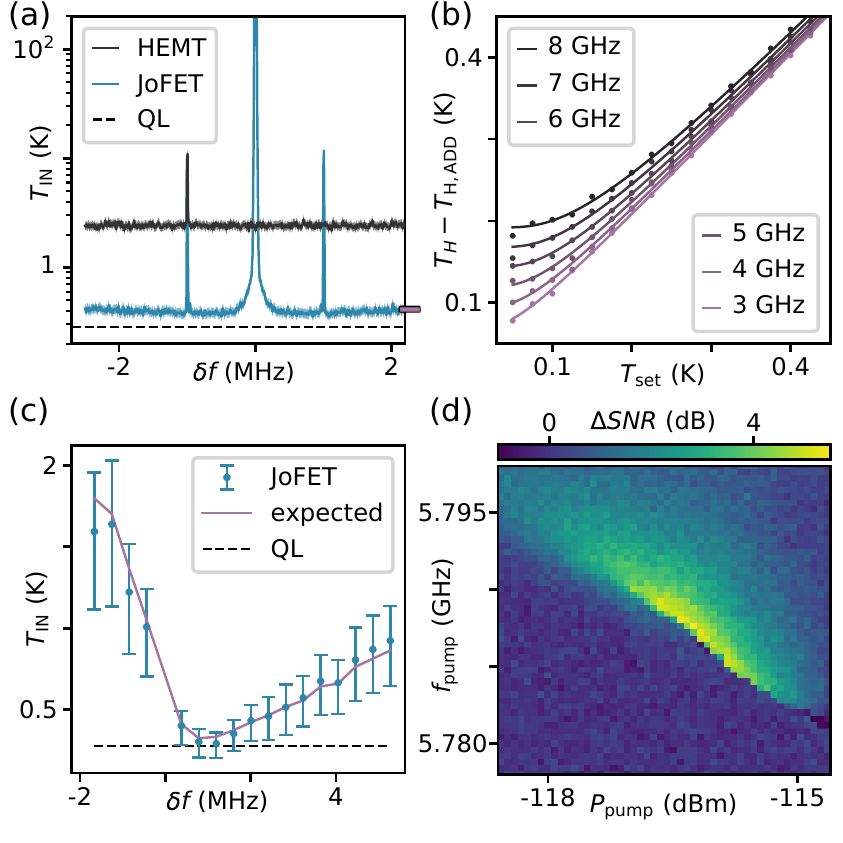}
		\caption{
			Noise performance.
			(a) Total noise temperature referred to JoFET input $T_\mathrm{IN}$, measured in the presence of a weak pilot tone, as a function of detuning frequency from pump $\delta f$.
			Pump off (HEMT, black) and pump on (JoFET, blue) shown, shown in comparison with the quantum limit at this frequency ($0.285~\mathrm{K}$, QL, dashed).
			$V_g=-2.35~\mathrm{V}$.
			Expected noise performance (Eq.~\ref{eq:s_t_expected}) is indicated by purple tick on the right-axis.
			In JoFET data, the blue-detuned idler and the zero-detuned pump are also visible.
			Bands represent propagated uncertainty from cavity parameters, insertion loss measurements, and HEMT noise calibration.
			Note that here HEMT data are referred to cavity input, not HEMT input.
			(b) Added noise of chain without JoFET, referred to HEMT input. Difference between total noise temperature $T_H$ and HEMT added noise $T_\mathrm{H,ADD}$ measured as a function of mixing chamber temperature setpoint $T_\mathrm{set}$. 
			Lines represent unity slope due to Johnson-Nyquist noise, corrected for the presence of vacuum fluctuations which dominate at low temperature and high frequency \cite{clerk_introduction_2010}.
			(c) $T_\mathrm{IN}$ measured as a function of pump-signal detuning $\delta f$ by varying pump frequency with signal frequency fixed (blue markers).
			Purple line is expected noise performance from Eq.~\ref{eq:s_t_expected}, and dashed line is the quantum limit at this frequency ($0.278~\mathrm{K}$).
			One datapoint with a detuning of $2.7~\mathrm{kHz}$ had to be removed from the dataset since the pump and signal could not be resolved individually.
			Error bars represent uncertainty propagated through from circuit parameters and HEMT noise calibration.
			(d) Relative improvement in signal to noise ratio ($\Delta \mathrm{SNR}$) measured as a function of pump frequency $f_\mathrm{pump}$ and pump power $P_\mathrm{pump}$: signal-pump detuning is fixed at $0.5~\mathrm{MHz}$, signal input power is fixed at $-153~\mathrm{dBm}$.
			$V_g = -2.55~\mathrm{V}$ for (c)-(d).
		}
		\label{fig:5}
\end{figure}

Quantum signals typically consist of few photons, making it crucial to achieve noise performance near the quantum limit.
To assess the noise performance of the JoFET amplifier, a weak pilot signal is measured first with a commercial HEMT amplifier, and then with parametric gain activated [Fig.~\ref{fig:5}(a)].
The JoFET amplifier dramatically improves the signal-to-noise ratio (SNR) of the pilot-signal measurement.
Our calibration procedure, described in the next paragraph and in App.~\ref{app:noise_referral}, gives an indication that the total input-referred noise of the JoFET amplifier and subsequent measurement chain approaches the limits placed by quantum mechanics of one photon from nondegenerate amplification \cite{caves_quantum_1982,clerk_introduction_2010}.
However, a definitive check of these results would require a calibrated noise source at device input.
The excess noise in the device is consistent with expectations based on resonator loss and finite gain [purple tick in Fig.~5(a)], as derived from input-output theory in App.~\ref{app:expected_noise}.

This noise measurement was carefully calibrated by varying the temperature of the mixing-chamber stage of the dilution refrigerator and measuring noise at various frequencies with the JoFET amplifier off [Fig.~\ref{fig:5}(b)].
In the high temperature limit, the output noise is linear in temperature, with an intercept that reflects the added noise of the chain referred to the mixing chamber plate, giving $T_\mathrm{H,MC}=1.61~\mathrm{K}$ at the JoFET operating frequency.
At low temperature input-referred noise saturates.
Calibrating over a wide range of frequencies reveals that noise saturation is pronounced only for high frequencies, consistent with the expected behavior for quantum fluctuations~\cite{clerk_introduction_2010}.
This gives evidence that quantum fluctuations are faithfully resolved by our measurement chain.
To find the noise referred to the JoFET input, $T_\mathrm{IN}$, we measured insertion loss of all components in-between the JoFET input and the mixing chamber plate, and used these values to refer the noise spectra to the device input (see Appendix \ref{app:noise_referral}).
This calibration procedure counts resonator and sample-holder insertion loss against the performance of the JoFET amplifier, resulting in a noise temperature that represents the JoFET added noise referred to its input, which is a suitable quantity for characterizing the JoFET as a standalone device.
It is not a measurement of total system efficiency, which would also need to include cable and circulator losses between a specified $50~\mathrm{\Omega}$ load and the JoFET.

Changing the pump frequency while keeping the signal frequency fixed at $5.7839~\mathrm{GHz}$ reveals that the noise performance is degraded if the pump frequency is away from the optimal operating region [Fig.~\ref{fig:5}(c)].
The change in noise performance with detuning matches the predictions of Eq.~\ref{eq:s_t_expected} with no free parameters [Fig.~\ref{fig:5}(c), purple line], indicating that the variation in noise performance is due to decreased gain.
The level of agreement between experiment and theory in Fig.~\ref{fig:5}(c) gauges the accuracy of our noise calibration method, although we again emphasize that a calibrated noise source is needed for a definitive noise measurement.
Note that this dataset was taken at a gate voltage with less resonator loss ($\kappa_\mathrm{ex}/\kappa=0.91$), so the expected noise performance is somewhat better than in Fig.~\ref{fig:5}(a), but the data do not resolve this difference due to large propagated uncertainties in Fig.~\ref{fig:5}(c).
By sweeping the pump frequency and the pump power with a fixed signal detuning, the optimum operating points of the amplifier can be extracted [Fig.~\ref{fig:5}(d)].
The operating points with best noise performance qualitatively correspond to the regions of highest gain identified in Fig.~\ref{fig:4}(c), as expected. 

\begin{figure}[bt]
	\centering
	  \includegraphics[scale=1]{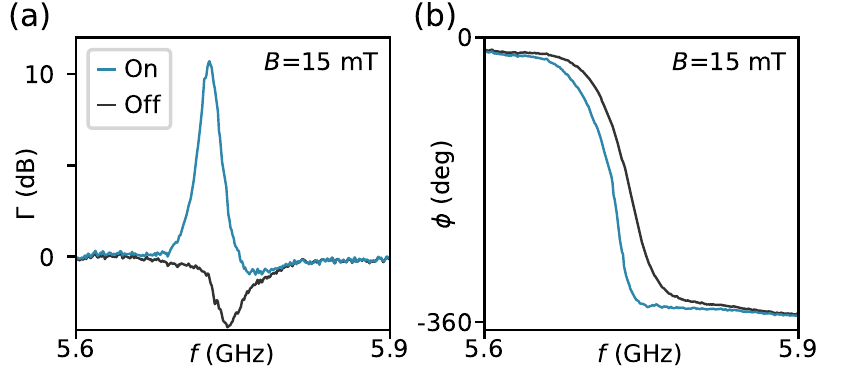}
		\caption{
		Magnetic field performance.
		Reflected magnitude $\Gamma$ and (a), and
		phase $\phi$ (b) with and without pump tone as a function of probe frequency $f$ in the presence of an in-plane magnetic field, $B=15~\mathrm{mT}$.
		}
		\label{fig:6}
\end{figure}

\section{Magnetic field operation}
A general technical advantage conferred by the Al-InAs hybrid platform is compatibility with large, parallel external magnetic fields \cite{pita-vidal_gate-tunable_2020,kringhoj_magnetic-field-compatible_2021,phan_detecting_2022}.
To explore if this advantage is realized in our particular device, we have steadily increased the parallel external magnetic field to $15~\mathrm{mT}$ while compensating for small field misalignments with a perpendicular magnetic coil.
Even with compensation, the resonator parameters evolve slightly in magnetic field, and a slight increase in loss is observed.
Applying a pump tone gives an optimized gain of $10~\mathrm{dB}$ [Fig.~\ref{fig:6}(a)-\ref{fig:6}(b)].
This demonstrates parametric amplification in a modest in-plane magnetic field, but far less than recently demonstrated high-field amplifiers~\cite{xu_magnetic_2022,khalifa_nonlinearity_2022,vine_in-situ_2022}.
Future experiments need to improve the resonator performance in magnetic field in order to allow a reliable noise measurement; we were unable to demonstrate quantum-limited operation in a magnetic field.
The most likely reason for this is the depinning of trapped flux by high pump powers.
This effect can be reduced by forming the resonator center pin from a field-compatible superconductor that connects to a smaller Al-InAs microstructure, as was done with recent field-compatible superconducting qubits \cite{pita-vidal_gate-tunable_2020,kringhoj_magnetic-field-compatible_2021}.

\section{Outlook}
Summarizing, we have demonstrated a high-performance JoFET amplifier, already obtaining performance comparable to early implementations in Al/AlOx material systems.
A number of techniques are available to further improve the performance of our device.
The participation factor of the superconductor-semiconductor heterostructure can be reduced drastically by working with a small mesa in the JoFET region.
This would likely decrease microwave losses, and allow the use of field-resilient superconductors like NbTi, which should allow operations in external magnetic fields of order 1 Tesla.
The number of JoFETs and the designed critical current can also be adapted to provide the target nonlinearity at milder gate voltages~\cite{eichler_controlling_2014}.
Device dissipation can also likely be improved by using a superconducting electrostatic gate with on-chip filters.
Finally, superconducting-semiconducting hybrid material systems are being actively developed, so material-level improvements can be expected.

Our work opens up the new, general direction of quantum-limited signal processing devices based on semiconductors, and can be expanded to devices such as modulators \cite{bergeal_analog_2010,chapman_single-sideband_2017} circulators \cite{chapman_widely_2017,chapman_design_2019} or signal generators~\cite{cassidy_demonstration_2017}.
It is interesting to note that the electrostatic gate can be viewed as a receiver for high-impedance electrical signals, suggesting applications such as electrometry with an integrated, quantum-limited amplifier.
Potential applications of electrometers include readout of spin qubits \cite{chatterjee_semiconductor_2021} and scanning-probe experiments \cite{yoo_scanning_1997,ilani_microscopic_2001}.
Given the fact that this device uses a narrow-gap III/V semiconductor, it is also interesting to consider applications in photodetection.

\textbf{Acknowledgements}
We thank Shyam Shankar for helpful feedback on the manuscript.
We gratefully acknowledge the support of the ISTA nanofabrication facility, MIBA machine shop and e-machine shop. 
NYU team acknowledges support from Army Research Office grant no. W911NF2110303. 

\textbf{Data availability}
Raw data for all plots in the main text and supplement will be included with the manuscript before publication.
Further data available upon reasonable request.

\appendix

\section{Sample preparation}
The JoFET amplifier was fabricated from Al-proximitized InAs quantum well which was grown on a semi-insulating, Fe counter-doped (100) InP wafer.
Patterning was performed with a Raith EBPG 5100 electron beam lithography system.
The Josephson weak link was made by etching a trench in the aluminum layer in the commercial etchant Transene D at $50~\mathrm{^\circ C}$ for $5~\mathrm{s}$.
The trench was designed to be $20~\mathrm{nm}$ long and $25~\mathrm{\mu m}$ wide.
Scanning electron microscopy (SEM) later showed that the trench width was about $50~\mathrm{nm}$.
The superconductor-semiconductor mesa forming the center pin of the CPW resonator was formed by masking with polymethyl methacrylate (PMMA) followed by a semiconductor wet etch in a mixture of $\mathrm{CH_{3}COOH:H_{2}O_{2}:H_{3}PO_{4}}$ for 150 seconds at room temperature.
The ground plane was constructed by evaporating Ti $5~\mathrm{nm}$, Nb $50~\mathrm{nm}$ following a short 1-min Ar ion milling at an accelerating voltage of $400~\mathrm{V}$, with an ion current of $21~\mathrm{mA}$ in a Plassys ultra-high vacuum (UHV) evaporator.
In the next step, the dielectric layer separating the gate and the junction was deposited with an Oxford atomic layer deposition (ALD) system running a thermal ALD alumina process at 150 degrees C in 150 cycles, which gave an approximated thickness of $12~\mathrm{nm}$.
Eventually, the gate that covers just the area of the Josephson weak link was created by evaporating Ti $8~\mathrm{nm}$ and Au $80~\mathrm{nm}$ at a tilt angle of 30 degrees with 5 RPM planetary rotation in a Plassys high vacuum (HV) evaporator.
Due to low adhesion of Al bond wire to Nb ground plane, co-deposited Au bond pads are used to enhance Al alloy formation during wire bonding, hence, chip-to-chip bond yield.
All lift-off and cleaning processes were performed in hot acetone at 50 degrees C and isopropanol.

\begin{figure}[bt]
	\centering
	  \includegraphics[scale=0.7]{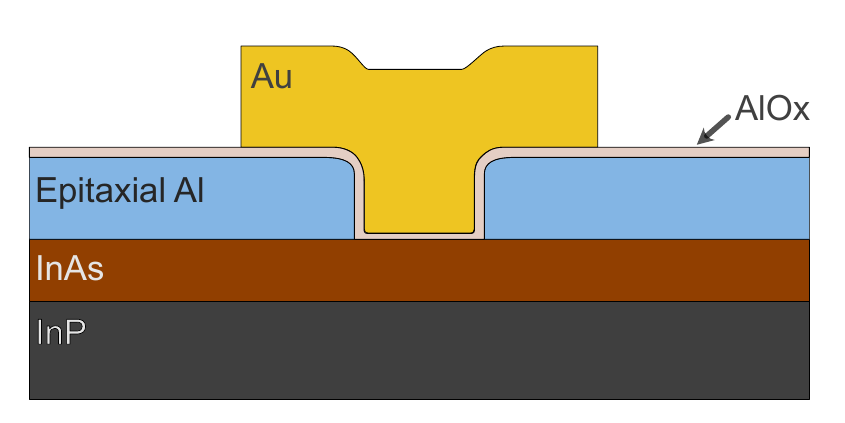}
		\caption{
		Cross section cartoon of the JoFET: (from bottom to top) the substrate is 500 $\mathrm{\mu m}$-thick InP(100), the quantum well is a heterostructure denoted InAs, the superconductor is a 10 $\mathrm{nm}$-thick layer of epitaxial Al, the dielectric that separates the Al and the gate is made with thermal ALD $\mathrm{AlO_{x}}$, the gate is made by evaporating $80~\mathrm{nm}$ of Au at a $30^\circ$ tilt angle. Cartoon dimensions are not to scale.
		}
	  \label{fig:7}
\end{figure}

The schematic in Fig.~\ref{fig:7} shows cross section of the JoFET.

\section{Measurements}
The sample was mounted on a copper bracket on a home made printed circuit board.
We used a vector network analyzer (VNA) model Keysight P9372A to measure scattering parameters.
A Rohde \& Schwarz (R\&S) signal generator model SGS100A was used to provide the pump tone.
When pumping at a fixed frequency, we used the VNA to provide a small probe signal and measure the gain profile from the reflected monotone at probe frequency.
A custom built low-noise voltage source provided the gate bias.
To achieve gain, we typically set the gate voltage to $-2.5~\mathrm{V}$ and swept the VNA power until the response became critical.
A pump tone from a R\&S signal generator was injected at a power of about $1~\mathrm{dB}$ below the approximated critical power and slightly above the critical frequency.
The probe power from the VNA was reduced to maintain the validity of the stiff pump condition.
We tuned up the JoFET amplifier by sweeping the pump frequency and power in the proximity of the nonlinear resonator's critical values and recording the maximal reflected amplitude at the VNA input.
Optimal amplification configurations were then found by fine sweeping around the regions where the maximal gains were highest, whose values were typically more than $20~\mathrm{dB}$.
To measure the noise, we recorded the power spectral density from the output of the JoFET amplifier by a ThinkRF R5550 hybrid spectrum analyzer.
In-plane magnetic field and compensation was applied by a vector magnet system of the Oxford Triton fridge and Mercury iPS power supply.

\section{RLC circuit model}
\label{app:rlc}
To model gate-dependence of $\kappa_\mathrm{ex}$ and $\kappa_\mathrm{i}$ in Fig.~\ref{fig:2} we find an effective lumped-element representation of the circuit, and then couple it to transmission lines following the procedure in Ref.~\cite{goeppl_coplanar_2008}.

Our circuit can be imagined as two resonators of length $l$ and inductance per unit length $L_l$, capacitance per unit length $C_l$, and attenuation constant $\alpha$ coupled through the parallel RL circuit.
For a large shunt resistance the inductance of the parallel combination is given by $L_J$.
In this limit the model therefore introduces dissipation without substantially altering the resonant frequency.
The resonant wave vector $k$ satisfies \cite{bourassa_josephson-junction-embedded_2012}
\begin{equation}
	2 \cot( k l ) = \frac{L_\mathrm{J}}{L_l l} k l.
\end{equation}

The effective capacitance is \cite{eichler_controlling_2014}
\begin{equation}
	C_\mathrm{eff} = C_l l \left(1 + \sinc( 2 k l ) \right).
\end{equation}
This expression arises from equating the capacitive energy in both resonators $2 \int_{-l}^0 dx C_l u(x)^2$ with an equivalent lumped-element energy $C_\mathrm{eff} u_i^2$, where $u(x) = -u_i \cos k_i (x+l)$ is a flux normal mode with wave vector $k_i$ \cite{bourassa_josephson-junction-embedded_2012,eichler_controlling_2014}.

We do not know of an explicit treatment of dissipation in our geometry available in the literature, so we seek an effective resistance in analogy with the effective capacitance discussed above.
The Rayleigh dissipation function, which entirely determines the effect of the resistance on circuit dynamics \cite{golstein_classical_2011}, is $\Delta \dot{u}(x)^2 / (2 R_J$), where $\Delta u$ is the flux drop across the resistor.
This has the physical interpretation of 1/2 of the dissipated electrical power.
Equating the dissipation function with an effective lumped-element dissipation rate $\dot{u}_i^2/(2 R_\mathrm{eff})$ suggests an effective resistance $R_\mathrm{eff} = R_J / (\Delta \overline{u})^2$, where $\Delta \overline{u} = 2 \cos(k l)$ is the normalized flux drop across the resistor.
Working in an effective lumped-element circuit model \cite{goeppl_coplanar_2008}, we then add $R_\mathrm{eff}$ in parallel with the resonator dissipation, finding expressions for the total dissipation and coupling rates,
\begin{eqnarray}
	\label{eq:ki}
	\kappa_i &=& \frac{\alpha l}{Z_0 C_\mathrm{eff}} + \frac{1}{R_\mathrm{eff} C_\mathrm{eff}} \\
	\label{eq:kex}
	\kappa_\mathrm{ex} &=& \frac{1}{R^* C_\mathrm{eff}},
\end{eqnarray}
where $R^*=(1 + \omega_k^2 C_k^2 Z_0^2 ) / (\omega_k^2 C_k^2 Z_0)$ is the effective parallel resistance from the measurement port with impedance $Z_0$ coupled with capacitance $C_k$.

To fit this model we use the fact that $kl = (\pi/2) f / f_0$ where $f_0$ is the bare resonant frequency of the CPW resonator when $L_J$ is zero.
As discussed in the main text, due to our uncertainty over how much inductance the junction contributes at large gate voltage we consider a range $f_0=6.00-6.45~\mathrm{GHz}$ and then extract $kl$ directly from the measured data. 
We fix the characteristic impedance of the resonator as $Z_0=50~\mathrm{\Omega} \cdot (f_\mathrm{geo}/f_0)$ based on the ratio between the designed geometric resonance frequency $f_\mathrm{geo}$ and the actual bare resonant frequency $f_0$, which physically originates from the kinetic inductance of the heterostructure \cite{phan_detecting_2022}.
Knowledge of $kl$, $f_0$, and $Z_0$ allows the effective capacitance and Josephson inductance to be calculated.
Eq.~\ref{eq:ki} is fit for $\alpha$ and $R_J$ in Fig.~\ref{fig:2}(a) and Eq.~(\ref{eq:kex}) is fit for $C_k$.

\section{Conversion from signal power to average intracavity photon number}
The average intra-cavity photon number $n$ is linearly dependent on the incident power $P_\mathrm{in}$ at the input port:
\begin{equation}
	n = \frac{1}{h f_s} \frac{4 \kappa_\mathrm{ex} P_\mathrm{in}}{(\kappa_\mathrm{ex}+\kappa_{i})^2 + 4 \Delta^2}.
\end{equation}
The input power is estimated based on the VNA power and the total attenuation $A$ of the setup.
This is used when expressing the theoretical Kerr nonlinearity as a frequency shift per input power (MHz/fW) in Fig.~\ref{fig:3}(d).

\section{Total attenuation estimate}
The total attenuation $A$ between the output of the VNA and the input of the device is calculated from noise spectrum from the measurement port with a signal tone at $-43~\mathrm{dBm}$ output power.
The noise floor is identified with input-referred noise temperature $T_\mathrm{H,MC}=1.61~\mathrm{K}$ from the HEMT calibration measurement shown in Fig.~\ref{fig:5}(b), combined with the resolution bandwidth of the signal analyzer $\mathrm{BW}=3.88~\mathrm{kHz}$, giving $P_\mathrm{noise}=10 \log( k_B T_N \mathrm{BW} /(1~\mathrm{mW})) = -160.6~\mathrm{dBm}$.
The signal is $6.3~\mathrm{dB}$ above the noise floor, so its input referred magnitude is $-153~\mathrm{dBm}$.
Accounting for the small $-0.84~\mathrm{dB}$ of loss from the (detuned) cavity then gives the total attenuation $A=-110~\mathrm{dB}$ quoted in the main text.
This value is compatible with expectations based on our installed attenuators and stainless steel cryogenic coaxial cables (estimated $-66~\mathrm{dB}$), and room temperature components ($-40~\mathrm{dB}$).

\section{Kerr nonlinearity extraction from power sweep}
\label{app:kerr}
The dependence of resonance frequency on input power was used to estimate the Kerr nonlinearity $K$ similar to \cite{castellanos-beltran_widely_2007}. 
Resonant frequency decreased linearly with increasing average intracavity photon number following the Hamiltonian \cite{eichler_controlling_2014}:
\begin{equation}
	H_\mathrm{JPA} = \hbar \left( \tilde{\omega}_0  + \frac{K}{2}  \left\langle A^\dagger A \right\rangle \right) A^{\dagger} A .
\end{equation}
The slope of the line in Fig.~\ref{fig:3}(c) measures $K/2$.

For low-transmission Josephson junctions, there is a fixed relationship between the Josephson energy $E_J$ and the Kerr nonlinearity \cite{bourassa_josephson-junction-embedded_2012,eichler_controlling_2014}
\begin{equation}
	\hbar K = -\frac{e^2}{2 C_\mathrm{eff}} \frac{L_\mathrm{eff}}{L_J} \Delta \overline{u}^4,
\end{equation}
where $L_\mathrm{eff}=(4 \pi^2 f_r^2 C_\mathrm{eff})^{-1}$ is the effective inductance and $\Delta \overline{u} = 2 \cos( k l )$ is the normalized flux drop across the junction.
We use this formula in Fig.~\ref{fig:3}(d).

\section{JoFET amplifier design}
CPW resonator was designed to have a characteristic impedance of $50~\mathrm{\Omega}$ with the center conductor width and gap of $25~\mathrm{\mu m}$ and $16~\mathrm{\mu m}$.
Resonator length was chosen so that the target resonant frequency of $6~\mathrm{GHz}$ would be achieved at base temperature.
To account for the kinetic inductance of the 10~\si{\nano\metre}-thick Al layer,  geometric resonant frequency was designed to be $7.2~\mathrm{GHz}$ so that result resonance would drop to $6~\mathrm{GHz}$ based on previous fabrication experience.
The kinetic inductance of the Al film fluctuates from device to device, presumably due to uncontrolled oxidation of the Al.
The JoFET was designed for a maximal critical current of $10~\mathrm{\mu A}$ which gives a zeroth order Kerr nonlinearity $K$ of $1.4~\mathrm{kHz}$ at zero bias and in this CPW configuration, such that the device can be gated into a regime appropriate for amplification \cite{eichler_controlling_2014}.
Designed dimensions were chosen based on our previous lithography tests and critical current measurements where a sheet critical current density of $0.38~\mathrm{\mu A/\mu m}$ was achieved.

\section{JoFET input noise referral}
\label{app:noise_referral}
\begin{figure}[bt]
	\centering
	  \includegraphics[scale=0.7]{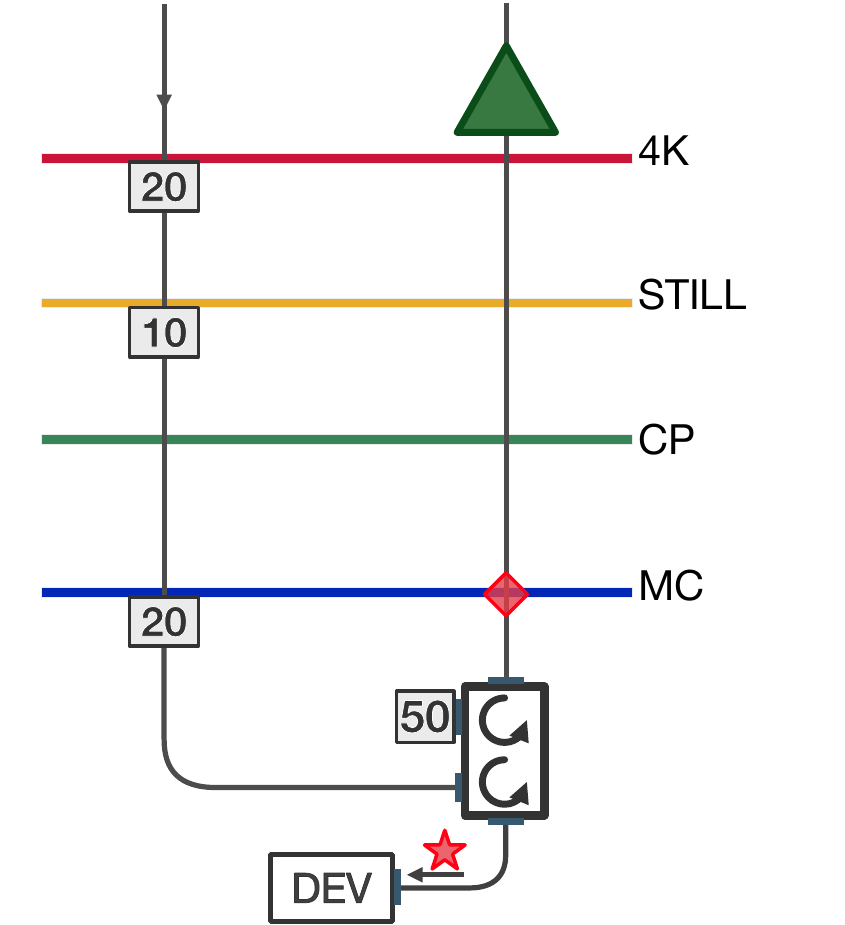}
		\caption{
		Schematic of insertion loss calibration.
		Noise referred to point $\diamond$ is referred to device input $\star$ by measuring the cold insertion loss of each component.
		Stainless steel coaxial cables and $50~\mathrm{dB}$ of additional attenuation thermalize signals on the input, and niobium coaxial cables are used between the 4K and MC stages on the readout chain.
		}
	  \label{fig:8}
\end{figure}
The temperature sweep in Fig.~\ref{fig:5}(b) is used to refer HEMT noise to the mixing chamber plate [point $\diamond$ in Fig.~\ref{fig:8}].
Noise is then referred to device input [point $\star$ in Fig.~\ref{fig:8}] by measuring the transmission $\eta$ of the components in the signal path: the circulators, coaxial cables, sample board, and device \cite{castellanos-beltran_widely_2007}.
We model this total loss as being from a beamsplitter with efficiency $\eta$ \cite{palomaki_entangling_2013}.

\begin{figure}[bt]
	\centering
	  \includegraphics[scale=0.95]{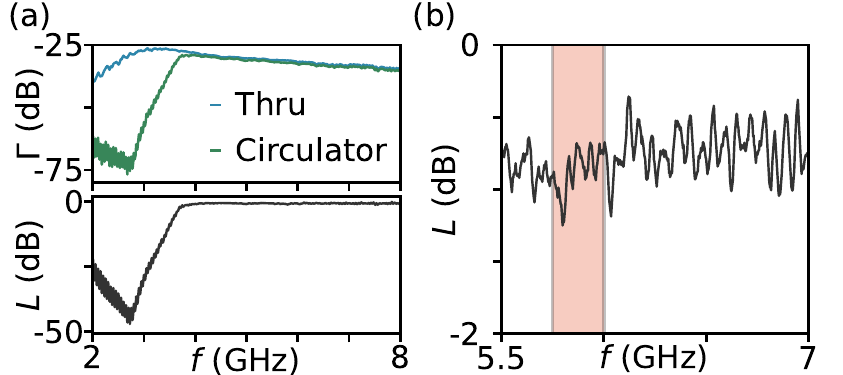}
		\caption{
		Calibration trace at base temperature in a dilution refrigerator.
		(a) Upper plot: Frequency-dependent transmission via a through cable (blue) and via the circulator and cables used for the measurements in the main text (green).
		Lower plot: frequency-dependent insertion loss $L$ calculated by the difference of the two traces in the upper plot.
		(b) Zoom on insertion loss over a relevant frequency region.
		To estimate the insertion loss in our experiment, we averaged over the frequency range $5.75-6~\mathrm{GHz}$ (shaded region), finding an insertion loss of $0.6\pm0.1~\mathrm{dB}$, where the uncertainty is from the standard deviation over the same range.
		}
	  \label{fig:9}
\end{figure}

The noise referred to $\diamond$, $S_\diamond$, is related to the real input noise at $\star$, $S_\mathrm{\star,in}$, according to $S_\diamond = \eta S_\mathrm{\star,in} + (1-\eta) V + T_\mathrm{H,MC}$, where the second term represents the introduction of vacuum noise $V$ by loss.
The total noise referred to point $\star$ is then 
\begin{equation}
\label{eq:input_referred}
S_\diamond/\eta = S_\mathrm{\star,in} + \frac{1-\eta}{\eta} V + \frac{1}{\eta} T_\mathrm{H,MC},
\end{equation}
which in the special case of vacuum-noise input gives $(T_\mathrm{H,MC}+V)/\eta$.

There are two important cases in the experiment: the measurement with the JoFET amplifier off [black line, Fig.~\ref{fig:5}(a)], and the measurement with the JoFET amplifier on [blue line, Fig.~\ref{fig:5}(a)]. 
In the off state, the pilot tone is not on resonance with the cavity, but still experiences a small loss $1-\eta_\mathrm{c,off}$ because of finite detuning.
In addition to cavity loss, there is an imperfect system transmission $\eta_s$ from the components between $\star$ and $\diamond$: the sample board, the circulators, and the coaxial cables.
This results in a net transmission in the off state of
\begin{equation}
	\label{eq:eta_off}
	\eta_\mathrm{off} = \eta_\mathrm{c,off} \eta_s.
\end{equation}
In the on state, there is a net gain $G$ from the device and the same system loss, so the net transmission in the on state is 
\begin{equation}
	\label{eq:eta_on}
	\eta_\mathrm{on} = G \eta_s.
\end{equation}
Note that, according to this definition, $G$ is the net gain of the amplifier including any cavity losses when it is on.

Input-referred noise is found by substituting into Eq.~(\ref{eq:input_referred}) $\eta=\eta_\mathrm{on}$ in the on case, and $\eta=\eta_\mathrm{off}$ in the off case.
Below we discuss how the three key parameters $\eta_s$, $\eta_\mathrm{c,off}$, and $G$ are measured.

We found $\eta_s$ by measuring the transmission through our cryostat at a base temperature $<0.05~\mathrm{K}$ with and without the cable/circulator combination [$0.6~\mathrm{dB}$ from Fig.~\ref{fig:9}], and also directly measured the sample holder insertion loss to be $0.2~\mathrm{dB}$ at liquid nitrogen temperatures.
These losses combine to give our system loss $\eta_s = 0.8$.

We found $\eta_{c,off}=0.87$ from measured scattering parameters in the same configuration, taking into account the detuning of the pilot from circuit resonance.
We emphasize that this number only includes a contribution from the circuit's off-resonant insertion loss, but not any system components such as the sample holder.

We found the gain $G$ by comparing the height of the pilot tone in the off state, $P_\mathrm{off}=\eta_\mathrm{off} P_\mathrm{\star,in}$, and in the on state, $P_\mathrm{on}=\eta_\mathrm{on} P_\mathrm{\star,in}$, for a fixed pilot power $P_\mathrm{\star,in}$.
Using Eq.~(\ref{eq:eta_off}-\ref{eq:eta_on}) we find the gain at the pilot frequency
\begin{equation}
	G = \eta_\mathrm{c,off} \frac{P_\mathrm{on}}{P_\mathrm{off}}.
\end{equation}
The frequency dependence of the gain is found from scaling a Lorentzian fit of the measured noise spectrum.

\section{Expected noise performance}
\label{app:expected_noise}

\begin{figure}[t]
	\centering
	  \includegraphics[scale=0.9]{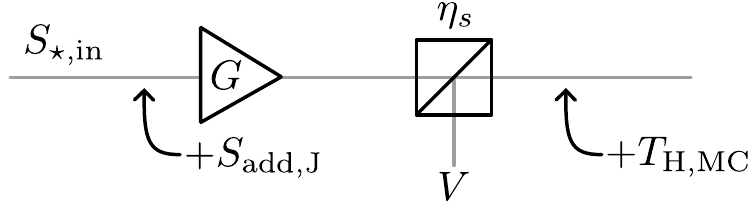}
		\caption{
		Noise model for estimating expected performance.
		Input noise $S_\mathrm{\star,in}$ is amplified by a JoFET amplifier with input-referred added noise $S_\mathrm{add,J}$ given by Eq.~\ref{eq:s_jpa_add} with on-resonant gain $g_S=\sqrt{G}$, followed by system losses modeled as a beamsplitter with efficiency $\eta_S$ and classical measurement chain with added noise temperature $T_\mathrm{H,MC}$.
		}
	  \label{fig:10}
\end{figure}

The quantum limit for phase-insensitive amplification is $2 V$ of input-referred added noise, where $V = h f / 2$ is the noise due to zero-point vacuum fluctuations \cite{caves_quantum_1982}.
Losses and finite gain cause deviations from this limit.
Here we determine the expected noise performance of our JoFET amplifier accounting for these effects.

We begin with the input-output relations for a Josephson parametric amplifier with input mode $a_{\mathrm{in},\Delta}$, output mode $a_{\mathrm{out},\Delta}$ at a detuning $\Delta$ from the pump frequency \cite{eichler_controlling_2014},
\begin{multline}
	\label{eq:jpa_io}
	a_{\mathrm{out},\Delta} = g_S a_{\mathrm{in},\Delta} + g_I a^\dagger_{\mathrm{in},-\Delta}+\\
	\sqrt{\frac{\kappa_i}{\kappa_\mathrm{ex}}} (g_S+1) b_{\mathrm{in},\Delta} +
	\sqrt{\frac{\kappa_i}{\kappa_\mathrm{ex}}} g_I b^\dagger_{\mathrm{in},-\Delta}.
\end{multline}
The bath mode $b_\mathrm{in,\Delta}$ is required by the fluctuation-dissipation relation in the presence of nonzero loss.
The complex signal gain $g_S$ is identically the scattering parameter $\Gamma$, and the idler gain $g_I$ ensures that $a_{\mathrm{out},\Delta}$ satisfies bosonic commutations relations \cite{clerk_introduction_2010}.
These commutation relations imply that
\begin{equation}
	\label{eq:bosonic_sumrule}
	(1+\frac{\kappa_i}{\kappa_\mathrm{ex}}) |g_I|^2 = |g_S|^2 -1 + \frac{\kappa_i}{\kappa_\mathrm{ex}} |g_S+1|^2.
\end{equation}
Assuming for convenience that the bath and idler modes are in their ground state with power spectral density $V$, the output spectral density of the JoFET amplifier $S_\mathrm{J}$ can be related to the input spectral density $S_\mathrm{\star,in}$ using Eq.~\ref{eq:jpa_io},
\begin{equation}
	S_\mathrm{J} = S_\mathrm{\star,in} |g_S|^2 + V \frac{\kappa_i}{\kappa_\mathrm{ex}} |g_S+1|^2 + V (1+\frac{\kappa_i}{\kappa_\mathrm{ex}})|g_I|^2.
\end{equation}
Eliminating $g_I$ using Eq.~\ref{eq:bosonic_sumrule} input referring by dividing out the gain gives $S_\mathrm{J}/|g_S|^2 = S_\mathrm{\star,in} + S_\mathrm{add,J}$, where the input-referred noise added by the JoFET amplifier is
\begin{equation}
	\label{eq:s_jpa_add}
	S_\mathrm{add,J} = V + 2 V \frac{\kappa_i}{\kappa_\mathrm{ex}} \frac{|g_S+1|^2}{|g_S|^2} - \frac{V}{|g_S|^2}.
\end{equation}
This expression represents the input-referred added noise of the JoFET amplifier accounting for both finite gain and resonator loss.
The first term is the quantum limit on added noise.
The second term gives the corrections due to resonator loss.
The final term gives a loss-independent correction for finite gain.
In the high-gain limit, Eq.~\ref{eq:s_jpa_add} reduces to a limiting value 
\begin{equation}
	\label{eq:s_jpa_approx}
	S_\mathrm{add,J} \approx V + \frac{2 \kappa_i}{\kappa_\mathrm{ex}} V,
\end{equation}	
which can be interpreted as the quantum limit plus unavoidable vacuum fluctuations arising from loss at the cavity input.
Equation~\ref{eq:s_jpa_approx} is sometimes referred to as the device quantum limit \cite{toth_dissipative_2017}.

As a check of Eq.~\ref{eq:s_jpa_add}, one can substitute $g_S = (\kappa_\mathrm{ex}-\kappa_i)/\kappa$, corresponding to the case where the amplifier is off and noise is measured at cavity resonance, and $S_\mathrm{\star,in}=~V$, corresponding to vacuum-noise input.
In this case one finds that the output noise is also vacuum, $S_\mathrm{JPA} = |g_S|^2 (V + S_\mathrm{add,JPA}) = V$, as expected.

To include the effect of system losses $\eta_S$ followed by our classical chain with temperature $T_\mathrm{HM,C}$, we combine Eq.~\ref{eq:s_jpa_add} with a simple beamsplitter noise model, as indicated in Fig.~\ref{fig:10}.
The total input-referred noise of the model at small detunings where the gain is real, $g_S=\sqrt{G}$, is 
\begin{equation}
\label{eq:s_t_expected}
S_\mathrm{\star,in}+S_\mathrm{add,J} + \frac{1-\eta_s}{\eta_s G} V + \frac{T_\mathrm{H,MC}}{\eta_s G}.
\end{equation}
Inserting $S_{\star,in}=V$ along with the measured device parameters and system loss from App.~\ref{app:noise_referral} gives a total expected input-referred noise of $0.41~\mathrm{K}$ for the data in Fig.~\ref{fig:5}(a) (indicated by purple tick right axis).
Eq.~\ref{eq:s_t_expected} is also used to calculate the expected noise performance in Fig.~\ref{fig:5}(c).

\section{Repeatability of calibration data}
\begin{figure}[t]
	\centering
	  \includegraphics[scale=0.9]{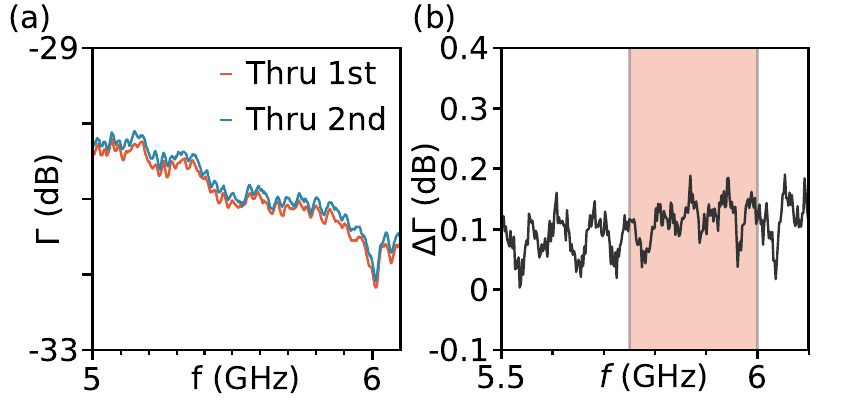}
		\caption{
		Repeatability of calibration data at base temperature in the dilution refrigerator.
		(a) Frequency-dependent transmission magnitude via a through cable from two different measurements.
		(b) Frequency-dependent difference of the transmission from the two traces in plot (a). The difference is smaller than $0.2~\mathrm{dB}$ in magnitude over the relevant (shaded) frequency range, and is on average $0.13~\mathrm{dB}$.
		}
	  \label{fig:11}
\end{figure}

The reproducibility of our circulator and cable insertion-loss measurements is key for the calibration described in App.~\ref{app:noise_referral}.
To check the stability of our method, more than one week after calibrating we removed the circulators and cables and repeated our through calibration measurement [Fig.~\ref{fig:11}(a)].
The two measurements differ by less than $0.2~\mathrm{dB}$ over the relevant frequency range, with an average difference of $0.13~\mathrm{dB}$.
Since our calibration occurred over a timespan of a few days, we interpret the $0.2~\mathrm{dB}$ value measured over one week as a conservative bound on drifts during our calibration procedure.
We therefore included an additional $0.2~\mathrm{dB}$ of uncertainty in the error bars in Figs.~\ref{fig:5}(a),\ref{fig:5}(c).

The calibrated components are inserted/removed via a bottom-loading exchange mechanism, which avoids having to thermally cycle the whole system.
We suspect that this feature greatly improves the reproducibility of our method.


\section{Summary of datasets}
Here we summarize the different datasets used in this work to facilitate their comparison.

\textit{Circuit characterization:} A single continuous dataset is used for Figs. \ref{fig:1}-\ref{fig:2}.
A single dataset is used to demonstrate circuit nonlinearities in Fig.~\ref{fig:3}(a)-\ref{fig:3}(c).
The nonlinearity is then characterized as a function of gate voltage in another dataset in Fig.~\ref{fig:3}(d).
We checked that the circuit resonant frequencies in Fig.~\ref{fig:3}(d) are compatible with those in Fig.~\ref{fig:1}(c).

\textit{Amplifier characterization:} A single dataset was used to show typical dependence of scattering parameters on pump power in Fig.~\ref{fig:4}(a)-\ref{fig:4}(b).
Consecutive datasets at a nearby ($50~\mathrm{mV}$ lower) gate voltage were then used to characterize the dependence of gain [Fig.~\ref{fig:4}(c)] and signal-to-noise ratio [Fig.~\ref{fig:5}(c),\ref{fig:5}(d)] on pump power and frequency.
The key amplifier metrics of input-referred noise [Fig.~\ref{fig:5}(a)] and compression point [Fig.~\ref{fig:4}(d)] were measured consecutively.
Noise was measured first, and compression point was measured at the same gate voltage approximately five hours later.
HEMT calibration in Fig.~\ref{fig:5}(b) was performed the following day.

\textit{Magnetic field data:} The two traces shown in magnetic field [Fig.~\ref{fig:6}(a),\ref{fig:6}(b)] were at identical gate voltage and taken under the same conditions.

%
	
\end{document}